\documentstyle[prl,multicol,epsf,aps]{revtex}
\draft

\title{Anomalous diffusion through coupling to a fractal environment:\\
 Microscopic derivation of the ``whip--back'' effect.} 
\author{Eric Lutz} 
\address{Max--Planck--Institut f\"ur Kernphysik,
  Postfach 103980, 69029 Heidelberg, Germany} 

\date{\today}

\def\openone{\leavevmode\hbox{\small1\kern-3.3pt\normalsize1}}

\newcommand{\la}{\langle}
\newcommand{\ra}{\rangle}

\newcommand{\be}{\begin{equation}}
\newcommand{\ee}{\end{equation}}
\newcommand{\bea}{\begin{eqnarray}}
\newcommand{\eea}{\end{eqnarray}}
\begin{document}

\twocolumn[\hsize\textwidth\columnwidth\hsize\csname@twocolumnfalse\endcsname
\maketitle

\begin{abstract}
Two models for quantum Brownian motion | the Oscillator Bath (OB)  model
and the Random Band--Matrix (RBM) model | are compared and a  relation between the
spectral density function $I(\omega)$ and the variance 
$\overline{{V_{ab}}^2}$ is established. The extension to a fractal
environment is then considered and the microscopic origin of anomalous diffusion
is discussed. In particular, it is shown that the asymptotic behavior of
the normalized velocity auto--correlation function (VACF) is entirely
determined by the band 
form factor. This allows for a microscopic derivation of the ``whip--back''
effect.
\end{abstract}
\vskip.5pc]
Anomalous diffusion has attracted considerable 
attention in the last decade (for a review see
\cite{bou90,shl94,pek98}) and numerous studies have been devoted to
its description, mostly at the phenomenological level. Among 
different approaches are fractional
diffusion equations \cite{sch89,met94,hil95}, L\'evy walks \cite{shl87},
fractional Fokker--Planck 
equations \cite{fog94,zas94,kol98,met99}, nonextensive
statistical mechanics \cite{zan95}, quantum 
L\'evy processes \cite{kus99} and generalized Langevin equations \cite{lil00}.
Anomalous  diffusion often results from the interaction with  a
complex, non--homogeneous background. Examples include 
 porous or disordered media \cite{bou90},
chaotic heat baths \cite{kus97} and also  fractal environments
\cite{gra87}. In this Letter we focus on a system  coupled to a
fractal heat bath and explore the microscopic origin of anomalous
diffusion with the help of the RBM model for quantum Brownian motion
introduced in Ref.~\cite{lut99}.  In order to make a direct 
comparison between the OB model \cite{gra88,wei99} and the RBM model,
and to  extent the 
latter  to the case of a fractal environment, we first
calculate the bath correlation function and  investigate the
validity of the Kubo--Martin--Schwinger (KMS) condition.  We then
relate  the 
VACF to the microscopic band form factor by employing the first
dissipation--fluctuation theorem and derive the ``whip--back'' effect
\cite{mur90,wan92}. 

We consider a quantum system $S$ weakly coupled to a heat bath
$B$. The coupling is taken linear in the position coordinate $x$ of
the system. The composite Hamiltonian  is given by 
\be
\label{eq1}
H = H_S\otimes \openone_B + \openone_S\otimes H_B  + x \otimes V\ , 
\ee
where $H_S$  ($H_B$) is the system (bath) Hamiltonian and $V$ is an
operator acting on the bath. It is assumed that at $t=0$, $S$
and $B$ are uncorrelated, so that the initial total density operator
 factorizes, 
$\hat{\rho}(0)=\hat{\rho}_S(0)\otimes\hat{\rho}_B(0)$, where
$\hat{\rho}_{S,B}(t)=\mbox{tr}_{B,S}[\hat{\rho}(t)]$. It is further
assumed that the bath is initially in thermal equilibrium at
temperature $T$, $ \hat{\rho}_B(0)= Z^{-1}\exp(-\beta H_B)$, with
$\beta = (kT)^{-1}$. 
 
In the standard approach to quantum dissipation \cite{gra88,wei99},
the  bath is 
modelled by an ensemble of independent harmonic oscillators (mass 
$m_i$, frequency $\omega_i$) and $V$ is assumed to be linear in the
positions $x_i$ of the oscillators, $V= \sum_i c_i x_i$, where
the $c_i$'s are  coupling constants. The coupling to the heat bath is
fully characterized by the spectral density function $I(\omega)$, which
is defined by
\be
\label{eq2}
I(\omega) = \pi \sum_i \frac{c_i^2}{2m_i\omega_i}\,
\delta(\omega-\omega_i)\ .
\ee   
On the other hand, in the random--matrix model  the
structure of the bath is not specified, in particular there are no
environmental oscillators, and  $V$ is a centered 
Gaussian random band--matrix. In this approach the coupling to the
bath is uniquely characterized by the second moment
$\overline{{V_{ab}}^2}$ of the random interaction.  The explicit form
used in Ref.\cite{lut99} reads 
\be
\label{eq3}
\overline{{V_{ab}}^2} = A_0 \left[ \rho(\varepsilon_a)
    \rho(\varepsilon_b)\right]^{-\frac{1}{2}}
    \exp\left[-\frac{(\varepsilon_a - \varepsilon_b)^2}{2\Delta^2}\right] \ .
\ee
Here $\varepsilon_a$'s denote the eigenenergies of the bath Hamiltonian
($H_B|a\ra = \varepsilon_a |a\ra$), $A_0$ is the strength of the
coupling, $\Delta$  the bandwidth and $\rho(\varepsilon)$ is the density
of states of the bath. Using the equivalence of the canonical and
microcanonical ensembles in the thermodynamical limit, the bath may be
parametrized as \cite{lut99}
\be
\label{eq4}
\hat{\rho}_B(0) = |a^*\rangle \langle a^*| \hspace{.6cm}
  \mbox{and} \hspace{.6cm} \rho(\varepsilon^*) = \rho_0\,
  e^{\beta \varepsilon^*} \ ,
\ee
where the state $|a\ra^*$ is defined by the temperature $T$ and
$\rho_0= Z/\sqrt{\pi kT^2 C_V}$ ($C_V$ being the heat capacity at
constant volume). 

In Ref.~\cite{lut99} we have demonstrated  that the averaged Markovian master
equation for  the system  obtained from  the RBM model is identical to 
the equation derived  starting from the OB
model. This shows that the two models, although being  completely
different in nature, are somehow related (at least at the level of the master
equation). It is our aim  to make 
this connection more precise and to  directly relate  the quantities
that define the two models, namely the spectral density function
(\ref{eq2}) and the variance (\ref{eq3}). This will be achieved by
evaluating the bath correlation function.

The bath correlation function is defined as
\be
\label{eq10}
K(t)= \la \widetilde{V}(t)\widetilde{V}(0) \ra_{B}\ ,
\ee
where $\widetilde{V}(t)=\exp(iH_B t)\, V \exp(-iH_B t)$ 
and $\la\dots\ra_B = \mbox{tr}_B[\hat{\rho}_B(0)\dots] $
denotes the thermal average. Performing the average over the
random--matrix ensemble in Eq.~(\ref{eq10}) leads to
\be
\label{eq11}
\overline{K}(t) = \int_{-\infty}^{+\infty} d\varepsilon_b\,
\rho(\varepsilon_b)\, \overline{{V_{ab}}^2}\, e^{i(\varepsilon_a-\varepsilon_b)t} \ .
\ee
The above equation  shows that $\overline{K}(t)$ is essentially the
Fourier transform of the  variance $ \overline{{V_{ab}}^2}$ with
respect to $\varepsilon_b$. On the
other hand, the
correlation function  can also  be written  in terms of 
 the spectral density $I(\omega)$ as
\bea
\label{eq12}
K(t)&=& \int_0^\infty \frac{d\omega}{\pi}
I(\omega)\left(\coth\left(\frac{\beta\omega}{2}\right) \cos(\omega t) - i
  \sin(\omega t) \right)\nonumber \\
&=& K'(t) + i K''(t) 
\eea
by using the identity of  $K(t)$ with the kernel of the influence
functional \cite{wei99}. 
We  see  from Eq.~(\ref{eq12}) that  the imaginary part $K''(t)$ of the
correlation function 
 is simply the Fourier sine
transform of the spectral density. Before we make use of this
observation to obtain the 
relation between $I(\omega)$ and $\overline{{V_{ab}}^2} $,
let us find the most general form of the variance allowed
by the laws of thermodynamics. More specifically, let us 
determine the conditions under which $\overline{K}(t)$ satisfies the
KMS condition,  $ \overline{K}(-t)=
\overline{K}(t-i\beta)$, which is known to define the thermal
equilibrium state of the bath \cite{spo78,thi83}. To this end, we consider
the following form of the variance ($ k>0 $), 
\be
\label{eq13}
\overline{{V_{ab}}^2} = A_0 \left[ \rho(\varepsilon_a)
    \rho(\varepsilon_b)\right]^{-k}
    \exp\left[-\frac{(\varepsilon_a -
        \varepsilon_b)^2}{2\Delta^2}\right]\ .
\ee
Substituting (\ref{eq13}) into Eq.~(\ref{eq11}), we obtain
\be
\label{eq14}
 \overline{K}(t)= \frac{A_0 \sqrt{2\pi}\, \Delta}{\rho(\varepsilon_a)^{2k-1}}
 \exp\left[-\frac{\Delta^2}{2}{\left(t+i
    (1- k)\beta \right)}^2\right]\ ,
\ee
where we have replaced the density of states $\rho(\varepsilon)$ by
the expression (\ref{eq4}).
We see from (\ref{eq14}) that the KMS condition is  only fulfilled
for  the value $k= 1/2$
(which is the form we have used so far) and that this is independent of
the actual form of the band. This can be understood by the 
following qualitative argument. Let us  write  $|\varepsilon_a - \varepsilon_b| = 
n d$ where $ d = \rho^{-1}$  is the mean level spacing and let us assume
that $nd \ll \Delta$. According to (\ref{eq13}), the intensity of the
coupling is 
approximately given by  $A_0 d^{2k}$. As a result, we observe  that  the
coupling is 
 much stronger for $k=0$ than for  $k=1/2$
 (recall that the spectrum
of $B$ is quasi-continuous which means that $d\sim 0$). This strong
coupling  then prevents the bath from being  in 
equilibrium. In the opposite limit 
 $k=1$,  the coupling is so weak that the imaginary part 
$\overline{K''}(t)$ vanishes. This implies that there is no
dissipation (see Eq.~(\ref{eq18})).  

The most general expression for the variance which respects the KMS
condition is hence of the form
\be
\label{eq15}
\overline{{V_{ab}}^2} = \left[ \rho(\varepsilon_a)
    \rho(\varepsilon_b)\right]^{-\frac{1}{2}} \mbox{f}(\omega_{ab}=
  \varepsilon_b-\varepsilon_a)\ , 
\ee
where we have introduced  the (symmetric) band form factor $
\mbox{f}(\omega)$. Note that the form (\ref{eq15}) ensures the passivity  
of the heat bath in accordance with the Second Law of
thermodynamics \cite{thi83} (for $k>1$ the damping coefficient defined
in Eq.~(\ref{eq19})
is  negative and the bath is clearly active) and also guarantees that
the bath remains in equilibrium at all times. Now, by comparing the imaginary
parts of Eqs.~(\ref{eq11}) and 
(\ref{eq12}), we arrive at the relation 
\be
\label{eq16}
I(\omega) = 2\pi \sinh\left(\frac{\beta\omega}{2}\right)
  \mbox{f}(\omega)\ .
\ee 
This equation provides a link between the two models. A
straightforward application of the above formula using the form
(\ref{eq3}) yields
\be
\label{eq17}
I(\omega) = 2\pi A_0 \sinh\left(\frac{\beta\omega}{2}\right)
\exp\left[-\frac{\omega^2}{2\Delta^2}\right]\ .
\ee
One may also  check that the power spectrum of 
$\overline{K'}(t)$ satisfies  the fluctuation--dissipation theorem
\cite{gra88,wei99} 
\be
\overline{K'}(\omega) = \coth \left(\frac{\beta\omega}{2}\right)
I(\omega) \ . 
\ee
At this point, it is useful to introduce the damping kernel  
\be
\label{eq18}
\gamma(t) = \frac{2}{M}\int_{-\infty}^t dt'\,\overline{K''}(t')\ .
\ee
In the limit $1\ll\Delta\ll kT$, (\ref{eq18}) is found to be equal to
\be 
\label{eq19}
\gamma(t) = 2\pi \frac{A_0\beta}{M}\, \delta(t) = 2\gamma \,\delta(t)\ .
\ee
With the help of Eq.~(\ref{eq19}),  Eq.~(\ref{eq17}) thus reduces to 
\be
I(\omega)\sim M\gamma\omega\ 
\ee
in the limit $\omega\rightarrow0$. 
Clearly, the variance (\ref{eq3}) defines Ohmic damping, as expected
from the results of Ref.~\cite{lut99}. Let us now turn to the
determination of  the band form factor in the
non--Ohmic case.

In the non--Ohmic regime, the spectral density of the bath
obeys $I(\omega)\sim \omega^\alpha, \alpha>0$, at low
frequencies. For 
$\alpha$ non--integer, the bath is called fractal \cite{leg87}. We
immediately see that
if the band form factor satisfies 
\be 
\label{eq200}
\mbox{f}(\omega)\sim g_\alpha \, |\omega|^{\alpha-1}\ ,
\ee
in the limit $\omega\rightarrow 0$, then, according to Eq.~(\ref{eq16}), 
\be 
\label{eq20}
I(\omega)\sim \pi\beta \,\omega \mbox{f}(\omega) \sim  \pi \beta \, g_\alpha \omega
^\alpha, \hspace{0.5cm}\omega>0 \ .
\ee
\begin{figure}[b!!]
\centerline{\epsfxsize=7.3cm
\epsfbox{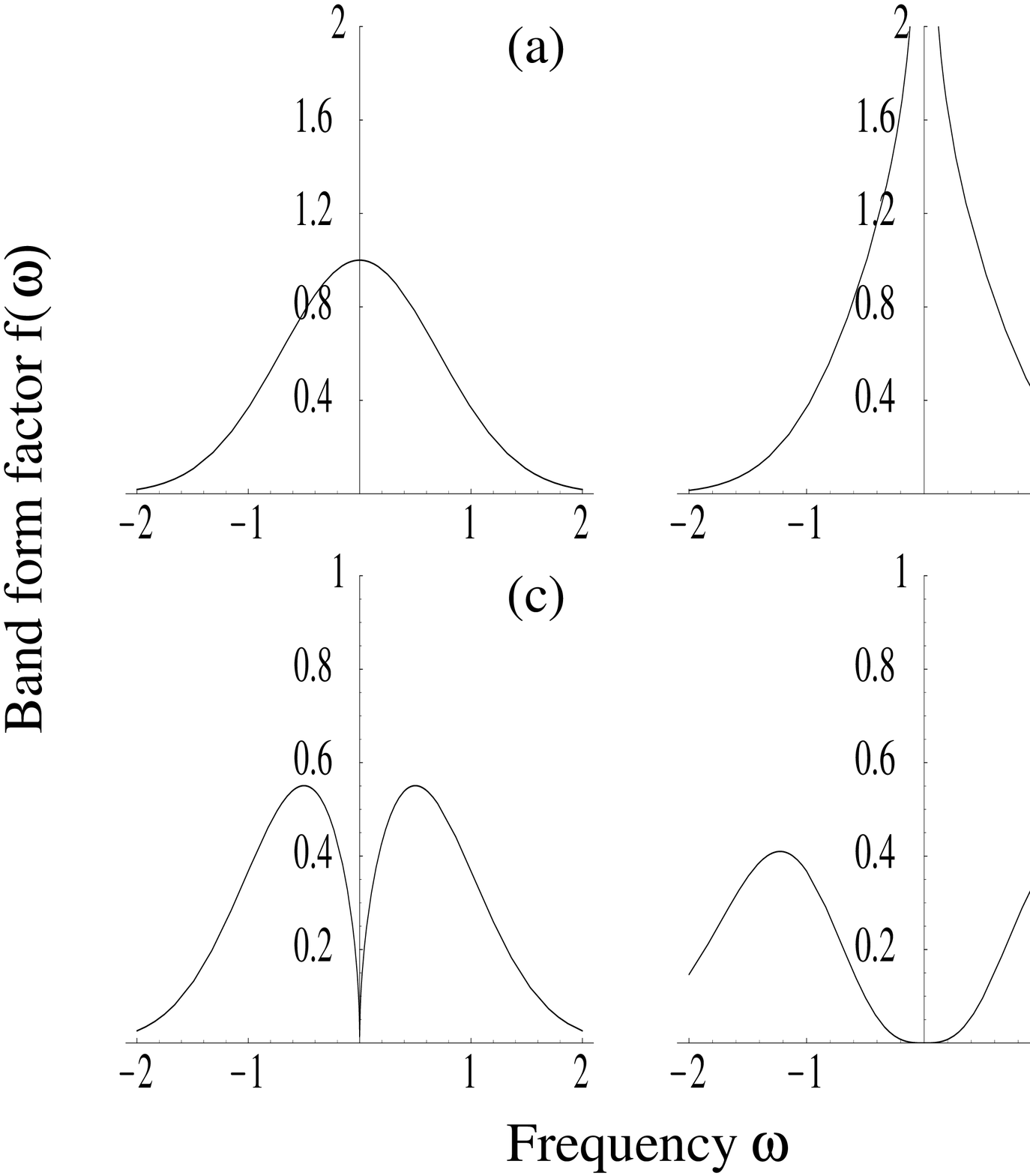}}
\vspace{.1cm}
\caption{Band form factor $\mbox{f}(\omega)=
  |\omega|^{\alpha-1}\,\exp[-\omega^2]$ for four  values 
   of  $\alpha $: (a) $ \alpha=1 $ (diffusive), (b) $
   \alpha=2/3 $ (subdiffusive), (c) $ \alpha=3/2 $ (superdiffusive),  (d) 
   $ \alpha=4 $ (ballistic).} 
\label{fig1}
\end{figure}
\noindent
We conclude that the damping regime is determined by the shape of
the band at  the origin  and that the  non-Ohmic (fractal)  regime is
the result  of a 
shape of type (\ref{eq200}). (Notice that the algebraic singularity
appearing 
in Eq.~(\ref{eq11}) for $\alpha<1$ can easily be regularized, since
the diagonal matrix elements $V_{aa}$ have no physical meaning, by
taking Hadamard's finite part $\int_{-\infty}^\infty x^{\alpha-1}
g(x)\, dx = \int_{-\infty}^\infty  
x^{\alpha-1} [g(x) - g(0)]\, dx $, see e.g.\cite{gel64}). 
Furthermore, it is known that the coupling to a
fractal environment gives rise  to anomalous diffusion at finite 
temperature \cite{gra87,gra88}. In the long--time limit, the
mean--square displacement 
$s(t) = \la(x(t)-x(0))^2\ra_t$ for a free damped particle grows like $
t^\alpha$ for $\alpha<2$ and like $ t^2$ for $\alpha>2$. This leads
to normal diffusion in the Ohmic case $\alpha=1$ and to
sub(super)diffusion in the case $\alpha<1$ ($1<\alpha<2$). For
$\alpha>2$, the growth of $s(t)$ is  ballistic. These different regimes 
can be qualitatively understood by looking at the behavior of the band
form factor at the origin (see Figure \ref{fig1}). Compared to the
diffusive case  $\alpha=1$, we observe that
subdiffusion is the consequence of the strong enhancement  
of small energy transfers which occurs for $\alpha<1$ (the absorption
by the bath
of a given amount of energy necessitates, on average, more steps in the
latter case: the bath becomes ``stiff'' and diffusion is slowed down), while
superdiffusion  results from the suppression of these small transitions 
for $1<\alpha<2$. The value $\alpha=2$ is
singled out in all anomalous  
diffusion phenomena. This  happens here too. When $\alpha>2$,
the derivative at the 
origin of the function $\omega^{\alpha-1}$ vanishes and
the form factor   $ \mbox{f}(\omega)$ shows a gap  around
$\omega=0$. As a result, the motion of the
free particle is dissipative only as long as its energy is larger than 
the value of the gap. Once its energy is smaller, the particle
behaves, on average, 
as if it were free (ballistic motion). The foregoing discussion can be 
made more quantitative and the microscopic origin of anomalous
diffusion more
transparent by evaluating the VACF of the
particle.

We shall now show that the long time behavior of the normalized  VACF 
$C_v(t)$ of the system  can be written in a
simple way in terms of the band form factor. Let us recall that the
canonical correlation function of the velocity operator $v$  is
defined as \cite{kub66}
\be
\label{eq21}
\la \tilde{v}(0);\tilde{v}(t)\ra = \beta^{-1} \int_0^\beta d\lambda\, 
\la\tilde{v}(-i\hbar \lambda )\tilde{v}(t)\ra\ .
\ee
In the limit $\hbar\rightarrow 0$, (\ref{eq21}) reduces to the classical 
correlation function $\la v(0) v(t)\ra$. The first
fluctuation--dissipation theorem \cite{kub66}
\be
\label{eq22}
C_v[z] = \Big(z+\gamma[z]\Big)^{-1}
\ee
then relates the Laplace transform $C_v[z]$ 
of  the
normalized VACF $C_v(t) = \la\tilde{v}(0);\tilde{v}(t)\ra 
/ \la\tilde{v}(0);\tilde{v}(0)\ra$ to  the Laplace transform  of the damping
kernel $\gamma(t)$, which is given by \cite{gra88}
\be
\label{eq23}
\gamma[z] = \frac{2}{M} \int_0^\infty \frac{d\omega}{\pi}
\frac{I(\omega)}{\omega}\frac{z}{z^2+\omega^2}\ .
\ee
We further note that in Eq.~(\ref{eq23})
\be
\label{eq24}
\lim_{z\rightarrow0}\,\frac{1}{\pi}\,
\frac{z}{z^2+\omega^2}= \delta(z) \ .
\ee 
This gives
\be
\label{eq25}
\lim_{z\rightarrow0}\, \gamma[z] = \frac{\pi\beta}{M}\,
\lim_{z\rightarrow0}\, \mbox{f}(z) \ ,
\ee
where we have used Eq.~(\ref{eq20}). Hence, by virtue of the final
value theorem \cite{hol66} (assuming that the limit of $C_v(t)$
exists), we can write
\be
\label{eq26}
\lim_{t\rightarrow\infty}C_v(t) = \lim_{z\rightarrow0} \Big(1+
\frac{\pi\beta}{M}\,z^{-1}\mbox{f}(z) \Big)^{-1}\ .
\ee
Eqs.~(\ref{eq25}) and (\ref{eq26}) show that the band form factor
governs both the asymptotic behavior of the damping kernel and of the
VACF. To be more precise, we shall assume
that the VACF is of the form $C_v(t)\sim t^\lambda$ for large times
and employ the following   
theorem on  the asymptotic properties of the Laplace transform of a
function $F(t)$ \cite{doe74} 
\bea\label{eq28}
\mbox{If }\hspace{1.1cm} F(t) &\sim& c \,t^\lambda, \hspace{2.1cm}t\rightarrow \infty \nonumber \\
\mbox{then } \hspace{0.5cm}zF[z] &\sim& c\, 
\Gamma(\lambda+1)z^{-\lambda},\hspace{0.5cm} z\rightarrow 0\ .
\eea   
This theorem  is a  consequence of 
\be
\label{eq29}
\int_0^\infty t^\lambda\, e^{-zt} dt = \Gamma(\lambda+1)\,
z^{-(\lambda+1)}\ .
\ee
In view of Eq.~(\ref{eq20}), we rewrite the right--hand side of
(\ref{eq26}) as
\be
\label{eq30}
z\,C_v[z]\sim \Big(1+\frac{\pi\beta}{M}\,
 g_\alpha z^{\alpha-2}\Big)^{-1},\hspace{0.5cm} z\rightarrow 0
\ee
Two cases have to be distinguished: $\alpha$  smaller or
larger than two. In the case  $ \alpha<2$, Eq.~(\ref{eq30})  reduces to
\be
\label{eq31}
z\,C_v[z]\sim \frac{M kT}{\pi g_\alpha} z^{-(\alpha-2)},\hspace{0.5cm}
z\rightarrow 0\ . 
\ee
By using  the theorem (\ref{eq28}), we then infer
\be
C_v(t) \sim \frac{M kT}{\pi g_\alpha}
\frac{t^{\alpha-2}}{\Gamma(\alpha-1)},\hspace{0.5cm} t\rightarrow
\infty\ .
\ee
For $\alpha < 1$ (subdiffusion), the normalized VACF has a negative
power--law tail (this can be seen by noting that $\Gamma(\alpha-1) =
\Gamma(\alpha)/(\alpha-1)$). This negative correlation leads to a
incessant   change of direction of  
the velocity of the
particle and  has  been named ``whip-back'' effect in
Ref.~\cite{mur90} (see also Ref.~\cite{wan92}) for this reason. This effect 
is responsible for   
the slower diffusion of the particle. In contrast, for   $\alpha
> 1$ (superdiffusion), the VACF possesses a  positive tail  which
means that the 
particle is more likely to move always in the same direction: this
results in a enhanced  diffusion. In the Ohmic case $\alpha=1$, the
VACF vanishes   for 
large times since $\Gamma(0)= \infty$ and the velocities are therefore 
uncorrelated. This is of course in agreement with the exact form of the VACF,
$C_v(t) = \exp(-\gamma t)$, which is obtained by taking the
inverse transform of Eq.~(\ref{eq22}) with
$\gamma[z]=\gamma$. On the other hand, in the case  $\alpha>2$, we
note that 
$z^{\alpha-2}$ tends to zero in the limit $z\rightarrow
0$. Eq.~(\ref{eq30}) can therefore be rewritten in the form
\be
\label{eq35}
z\,C_v[z]\sim 1,\hspace{0.5cm} z\rightarrow 0 \ ,
\ee
and we have  accordingly
\be
\label{eq36}
 C_v(t) \sim 1,\hspace{0.5cm} t\rightarrow \infty \ .
\ee
The VACF (\ref{eq36}) corresponds to  ballistic motion.

In conclusion, we have investigated the validity of the KMS condition
for the RBM model by calculating the bath correlation
function. We have found that the most general form of the
variance compatible with the Second Law of thermodynamics is given by
Eq.~(\ref{eq15}). This form guarantees the passivity of the heat bath.
We have further derived a relation (Eq.~(\ref{eq16})) between the
spectral density of the  
OB model and the band form factor and extended the latter to a fractal
environment. We have then discussed the microscopic origin of
anomalous diffusion; first qualitatively by examining the form factor
at the origin, and second, quantitatively, by expressing the
asymptotic behavior of the normalized VACF in terms of the band form
factor. This resulted in the derivation of the ``whip--back' effect.

We wish to thank H. Grabert for having suggested the direct comparison
between the RBM and OB models and H.A. Weidenm\"uller for useful discussions.

\end{document}